\documentclass[twocolumn,epjc3]{svjour3} 

\pdfoutput=1

\usepackage[T1]{fontenc} 
\usepackage[utf8]{inputenc}
\usepackage{booktabs}
\usepackage{amsmath}
\usepackage{amssymb}
\usepackage{graphicx}
\usepackage{multirow}

\usepackage{xcolor}

\newcommand{\preprint}[1]{\gdef\varpreprint{#1}}

\journalname{Eur. Phys. J. C}
\begin{document}

\title{Neutrino mass models: New classification and model-independent upper limits on their scale}

\author{Juan Herrero-Garc\'{\i}a\thanksref{SISSA} \and Michael A.~Schmidt\thanksref{UNSW}}
\institute{SISSA/INFN, Via Bonomea 265, I-34136 Trieste, Italy, \label{SISSA}
	\email{jherrero@sissa.it}
\and 
School of Physics, The University of New South Wales, Sydney, NSW 2052, Australia,  \label{UNSW}
\email{m.schmidt@unsw.edu.au}
}
\preprint{SISSA  07/2019/FISI}
\maketitle
\begin{abstract}
	We propose a model-independent framework to classify and study neutrino mass models and their phenomenology. The idea is to introduce one particle beyond the Standard Model which couples to leptons and carries lepton number together with an 
	operator which violates lepton number by two units and contains this particle. This allows to study processes which do not violate lepton number, while still working with an effective field theory. The 
	contribution to neutrino masses translates to a robust upper bound on the mass of the new particle. We compare it to the stronger but less robust upper bound from Higgs naturalness and discuss several lower bounds. Our framework allows to classify neutrino mass models in \emph{just} 20 categories, further reduced to 14 once nucleon decay limits are taken into account, and \emph{possibly} to 9 if also Higgs naturalness considerations and direct searches are considered.
\end{abstract}

\section{Introduction\label{sec:intro}}

Neutrino oscillation experiments established the need for massive neutrinos and large mixings in the lepton sector. At the same time tritium beta decay experiments, cosmology and experiments searching for neutrinoless double beta decay put strong constraints on the absolute scale of neutrino mass. Despite tremendous progress in neutrino physics in recent years, the origin of neutrino mass remains a mystery. 

An elegant explanation of small neutrino masses is obtained by linking their smallness to the breaking by two units of lepton number ($\mathcal{L}$), the number of leptons minus antileptons, at a high scale $\Lambda$.
This leads to a plethora of explicit models such as
the tree-level seesaw
models~\cite{Minkowski:1977sc,Yanagida:1979as,GellMann:1980vs,Mohapatra:1979ia,Glashow:1979nm,%
Magg:1980ut,Schechter:1980gr,Lazarides:1980nt,Wetterich:1981bx,Mohapatra:1980yp,%
Foot:1988aq}
and models at loop level (see Refs.~\cite{Zee:1980ai,Cheng:1980qt,Zee:1985rj,Zee:1985id,Babu:1988ki,Ma:2009dk,Boucenna:2014zba,Cai:2017jrq} for the
first one- and two-loop models and recent reviews).
There are also several systematic studies of neutrino mass
generation~\cite{Bonnet:2012kz,Farzan:2012ev,Sierra:2014rxa,Weinberg:1979sa,Babu:2001ex,deGouvea:2007qla,delAguila:2012nu,Angel:2012ug,Angel:2013hla,deGouvea:2014lva,Cai:2014kra,Cepedello:2017eqf,Cepedello:2018rfh,Anamiati:2018cuq},
in particular studies of Majorana neutrino mass generation in terms of effective
operators that break lepton number by two units ($\Delta\mathcal{L}=2$) ~\cite{Weinberg:1979sa,Babu:2001ex,deGouvea:2007qla,delAguila:2012nu,Angel:2012ug,Angel:2013hla,deGouvea:2014lva,Cai:2014kra},
which provide an efficient way to study neutrino mass generation,
but do not allow to study other phenomenology such as lepton flavour violating processes
or searches at colliders.

Here we propose a hybrid approach in order to use the best of both schemes. 
It is based on the following premises: 
\begin{enumerate}
\item In any model of Majorana neutrino masses there is at least one new particle of mass $M$ which directly couples to leptons and carries lepton number (and in some cases also baryon number $\mathcal{B}$). We assume that this is the lightest beyond the Standard Model (SM) particle involved in the generation of neutrino masses. 
\item Following the common lore in quantum field theory that \emph{everything not forbidden is mandatory}, lepton number is violated by two units ($\Delta \mathcal{L}=2$) via operators\footnote{More precisely, the combination of both interactions (given in columns 2 and 3 in Tab.~\ref{tab:particles}) violates lepton number by two units. While the induced $\Delta L=2$ SM operator is odd-dimensional~\cite{deGouvea:2014lva,Kobach:2016ami}, the $\Delta L=2$ operator with one copy of the new particle may be even-dimensional depending on the new particle and its interactions, in particular this may be the case if the new state is fermionic.} which contain the new particle. 
\item Neutrino masses are generated from the $\Delta \mathcal{L}=2$ interactions of the new particle. We assume that this contribution dominates and generates the scale of neutrino mass, $m_\nu \gtrsim \sqrt{\Delta m_{\rm atm}^2}\simeq 0.05$ eV. The latter can be estimated~\cite{deGouvea:2007qla} and recast into a conservative upper bound on $M$.\footnote{Similarly the upper bound on neutrino masses can be translated in a lower limit on $\Lambda$ (but not on $M$) which is of similar size.}
\end{enumerate}
Most models in fact require to add more than one particle.\footnote{Thus this approach does not include models where the new particles are charged under extra symmetries beyond the SM gauge group. One example are models where the new particles couple in pairs to the SM leptons, like in the Scotogenic model and its Generalised versions~\cite{Ma:2006km,Ma:2013yga,Hagedorn:2018spx}. In this cases the new states have new global symmetries (discrete or continuous), and a DM candidate is present. These type of scenarios will be studied in future work.}. In our approach the effect of the additional particles is encoded in the $\Delta \mathcal{L}=2$ operators. In order to derive upper bounds, we only consider the lowest-dimensional and simplest $\Delta \mathcal{L}=2$ operator. We further assume order one couplings for all new interactions, and that third generation SM fermions dominate, which are the most conservative options.
In contrast to approaches based on effective operators alone, the introduction of the new particle enables to study processes which do not violate lepton number and their constraints on neutrino mass generation without going to explicit models. There are in total only 20 different categories, listed in Tab.~\ref{tab:particles}, which describe the theory space which is consistent with the first premise.
In the following we will first discuss upper bounds on the mass of the new states, and briefly several lower bounds. A more detailed discussion of the latter is left for future work.

\section{Upper bounds\label{sec:upper}}

For Majorana neutrinos 
the dominant contribution to neutrino masses
generally originates from the unique dimension 5 operator
$\mathcal{O}_1\equiv LH LH$, the so-called Weinberg
operator~\cite{Weinberg:1979sa}, where $L$ ($H$) is the SM lepton (Higgs) doublet. After electroweak symmetry breaking it leads to $m_\nu
\simeq c_1\, v^2/\Lambda$, with
$\langle H\rangle=(0,v)^T$, $v\simeq 174$ GeV and $c_1/\Lambda$ the Wilson coefficient of $\mathcal{O}_1$. The smallness of
neutrino mass is generally linked to the hierarchy $v\ll \Lambda$, known as the seesaw mechanism~\cite{Minkowski:1977sc,Yanagida:1979as,GellMann:1980vs,Mohapatra:1979ia,Glashow:1979nm,Magg:1980ut,Schechter:1980gr,Lazarides:1980nt,Wetterich:1981bx,Mohapatra:1980yp,Foot:1988aq}. For $c_1 \sim \mathcal{O}(1)$ the scale $\Lambda$ has to be sufficiently small, $\Lambda\lesssim 6\times 10^{14}$ GeV, so that $m_\nu \gtrsim 0.05\, \mathrm{eV}$.
Some models may feature an additional suppression encoded in the parameter
$\epsilon$. It may be due an almost conserved lepton number like
in the type-II seesaw model
($\epsilon=\mu/m_\Delta$)~\cite{Magg:1980ut,Schechter:1980gr,Cheng:1980qt,Lazarides:1980nt,Wetterich:1981bx,Mohapatra:1980yp}, inverse seesaw scenarios ($\epsilon=\mu/m_R$)~\cite{Mohapatra:1986aw,Mohapatra:1986bd}, or the (Generalised) Scotogenic model ($\epsilon=\lambda_5$)~\cite{Ma:2006km,Ma:2013yga,Hagedorn:2018spx}. In all these cases lepton number is restored in the limit $\epsilon\to 0$. 
Similarly, in models where the Weinberg operator is absent but $\mathcal{O}_1^{\prime n}\equiv LH LH (H^\dagger H)^n$ is generated, neutrino masses are suppressed by $(v^2/\Lambda^2)^n$~\cite{Bonnet:2009ej}. 
Finally neutrinos may be massless at tree level and only be generated at loop level.
Hence, 
it is better to parameterise neutrino mass by 
\begin{equation} \label{eq:mnueq}
m_\nu \simeq \frac{c_{\rm R} v^2}{(16\pi^2)^{\ell}\Lambda}\,, \quad \text{with}\quad c_{\rm R} \simeq  \prod_i g_i\,\times\, \epsilon\, \times \,\left(\frac{v^2}{\Lambda^2}\right)^n\,,
\end{equation}
where $i$ runs over the couplings $g_i$ and $\ell$ is the loop order at which neutrino mass is generated.
The couplings $g_i$ are subject to perturbativity 
constraints, which naively demands them to be at most order one. 
For low-scale models rare processes typically constrain the couplings to be even smaller, naively
$g_i\lesssim\mathcal{O}(0.1)$. The number of couplings increases with the loop
order. A conservative estimate yields that there are at least $2\ell$
couplings in an $\ell$ loop diagram and thus 
$\Lambda \lesssim 10^{12}\, (10^{10})\, [10^{8}]$ GeV using $\mathcal{O}(1)$ couplings for neutrino masses generated at one
(two) [three] loop order. Neutrino mass generation at higher loop order is 
thus theoretically disfavoured. These simple estimates
however do not allow to distinguish further between different 
models and thus it is desirable to go beyond.

\begin{table*}[tbhp!]\centering
\begin{tabular}{   c c  c   c c c c c  }\toprule 
	Particle & $\Delta \mathcal{L}=0$ & $|\Delta \mathcal{L}|=2$
		 & BL & $\ell$&$m_\nu$ & Upper bound \\ \midrule 
	$\bar N\sim(1,1,0)_F^{-1,0}$&$y\,\bar N H L $ & $M\,\bar N\bar N $ 
			       & $\mathcal{O}_1$&0&$\frac{y^2\,v^2}{M}$ & $M\lesssim 10^{15}$ GeV\\ 
$\Delta \sim(1,3,1)_S^{-2,0}$ & $y\,L\Delta L$& $\mu\, H \Delta^\dagger H$
			      & $\mathcal{O}_1$&0&$\frac{y\, \mu\,v^2}{M^2}$ & $M \lesssim 10^{15}$ GeV\\ 
$\bar\Sigma_0\sim(1,3,0)_F^{-1,0}$&$y\,\bar \Sigma_0 L H$ & $M\, \bar\Sigma_0\bar\Sigma_0$ 
			    & $\mathcal{O}_1$&0&$\frac{y^2\,v^2}{M}$ & $M\lesssim 10^{15}$ GeV \\

\multirow{2}{*}{$L_1\sim(1,2,-1/2)_{F}^{1,0}$}
&  $m \,\bar L_1 L$ & $\frac{c\,}{\Lambda}  L_1 H L H$ 
					 & $\mathcal {O}_1$&0&$\frac{c\, m}{M}\,\frac{v^2}{\Lambda}$ c& $M\lesssim 10^{15}$ GeV  \\ 

					 &$y\,H^\dagger \overline{e} L_1$ & $\frac{c\,}{\Lambda^2} \bar L_1 \bar u \bar d^\dagger L^\dagger$ 
				     & $\mathcal {O}^\dagger_{8}$&2&$\frac{c\,y y_u\,y_d\,y_l}{(4\pi)^4}\,\frac{v^2}{\Lambda}$  & $M\lesssim 10^7$ GeV  \\

\midrule
	$h\sim(1,1,1)_S^{-2,0}$&$y\,LL h$& $\frac{c}{\Lambda} h^\dagger \overline{e} LH$
			       & $\mathcal{O}_2$&1&$\frac{c\,y\,y_l}{(4\pi)^2}\,\frac{v^2}{\Lambda} $ & $M \lesssim 10^{10}$ GeV  \\ 

	$k\sim(1,1,2)_S^{-2,0}$&$y\,\bar e^\dagger \bar e^\dagger k$ & $\frac{c}{\Lambda^3} k^\dagger  L^\dagger L^\dagger L^\dagger L^\dagger $ 
			       & $\mathcal {O}^\dagger_{9}$&2&$\frac{c\,y\, y_l^2}{(4\pi)^4}\,\frac{v^2}{\Lambda} $ & $M\lesssim 10^6$ GeV\\ 
%
%
\multirow{2}{*}{	$\bar E\sim(1,1,1)_{F}^{-1,0}$}&$y\, \bar E L  H^\dagger $ 			& $\frac{c\,}{\Lambda^4} L E H Q^\dagger \bar u^\dagger H$ 
				   & $\mathcal{O}_6$ 
				   &2& $\frac{c\,y\,y_u}{(4\pi)^4} \frac{v^2}{\Lambda} $ 
				   & $M\lesssim 10^{10}$ GeV \\
				   &  $m \,\bar e E$ & $\frac{c\,}{\Lambda^3}  \bar E L L L H$  
				   & $\mathcal {O} _2$&1&$\frac{c\, m}{M}\,\frac{y_l}{(4\pi)^2}\,\frac{v^2}{\Lambda}$  & $M\lesssim 10^{10}$ GeV  \\ 

	$\bar \Sigma_1\sim(1,3,1)_{F}^{-1,0}$&$y\, H^\dagger \bar \Sigma_1 L$ 				      & $\frac{c}{\Lambda^2} LH  H \Sigma_1 H$     
				   & $\mathcal{O}^{\prime1}_1$&1& $\frac{c\,y}{(4\pi)^2} \frac{v^2}{\Lambda}$  & $M\lesssim 10^{12}$ GeV \\ 

	$L_2\sim(1,2,-3/2)_{F}^{1,0}$&$y\,H\overline{e} L_2$ & $\frac{c\,}{\Lambda^2} \bar L_2 L LL$
				     & $\mathcal {O}_{2}$&1&$\frac{c\,y\,y_l}{(4\pi)^2}\,\frac{v^2}{\Lambda}$  & $M\lesssim 10^{11}$ GeV  \\ \midrule

$X_2\sim (1,2,3/2)_V^{-2,0}$ & $ y\, \bar e^\dagger \bar\sigma^\mu L X_{2\mu}$ &  $ \frac{c}{\Lambda}\, \bar u^\dagger \bar\sigma^\mu \bar d X_{2\mu}^\dagger H$ 
			   &$\mathcal{O}_8$ 
			   &2& $\frac{c y\,y_u y_d y_e}{(4\pi)^4} \frac{v^2}{\Lambda}$ & $M\lesssim 10^7$ GeV
\\
\midrule

$\tilde R_2\sim(3,2,1/6)_S^{-1,1}$ &$y\, \overline{d} L \tilde R_2$ & $\frac{c}{\Lambda} \tilde R_2^\dagger Q L H  $ 
				   & $\mathcal {O}_{3_b}$
				   &1&$\frac{c\,y\,y_d}{(4\pi)^2}\,\frac{v^2}{\Lambda}$  & $M\lesssim 10^{11}$ GeV\\ 
\multirow{2}{*}{$R_2\sim(3,2,7/6)_S^{-1,1}$}& $y\,\bar e^\dagger  Q^\dagger R_2$ & $\frac{c}{\Lambda^3}R_2^\dagger L^\dagger L^\dagger L^\dagger\bar d^\dagger$ 
			   & $\mathcal {O}^\dagger_{10}$&2&$\frac{c\,y\,y_d\,y_l}{(4\pi)^4}\,\frac{v^2}{\Lambda}$ & $M\lesssim 10^7$ GeV\\ 

&$y\, \bar u L R_2$ & $\frac{c}{\Lambda^3}R_2^\dagger L^\dagger L^\dagger L^\dagger\bar d^\dagger$
&$\mathcal{O}^\dagger_{15}$ 
&3&$\frac{c\,y\,y_d\,y_u\,g^2}{2(4\pi)^6}\,\frac{v^2}{\Lambda}$ & $M\lesssim 10^6$ GeV\\

\multirow{2}{*}{$S_1\sim(\overline{3},1,1/3)_S^{-1,-1}$} & $y\, LQ S_1$ &  $\frac{c}{\Lambda}S_1^\dagger L H\overline{d}$ 
							 & $\mathcal {O}_{3_b}$&1&$\frac{c\,y\,y_d}{(4\pi)^2}\,\frac{v^2}{\Lambda}$  & $M\lesssim 10^{11}$ GeV\\

							 &$y\,\bar u^\dagger \bar e^\dagger S_1$   & $ \frac{c}{\Lambda} S_1^\dagger  L H \bar d$
							 & $\mathcal{O}_8$ 
							 & 2&$ \frac{ c\, y\,y_l\,y_u\,y_d }{(4\pi)^4} \frac{v^2}{\Lambda}$ & $M\lesssim 10^7$ GeV \\

$S_3\sim(\overline{3},3,1/3)_S^{-1,-1}$&$y\,L S_3 Q$ & $\frac{c}{\Lambda}\overline{d}\,L S_3^\dagger H$
				       & $\mathcal {O}_{3_b}$&1&$\frac{c\,y\,y_d}{(4\pi)^2}\,\frac{v^2}{\Lambda}$  & $M\lesssim 10^{11}$ GeV  \\ 

$\tilde S_1\sim(\bar 3,1,4/3)_S^{-1,-1}$&$y\,\bar e^\dagger \bar d^\dagger \tilde S_1$ & $\frac{c}{\Lambda^3} \tilde S_1^\dagger L^\dagger L^\dagger L^\dagger Q^\dagger$ 
				       & $\mathcal {O}^\dagger_{10}$&2&$\frac{c\,y\,y_d\,y_l}{(4\pi)^4}\,\frac{v^2}{\Lambda}$  & $M\lesssim 10^7$ GeV  \\
\midrule

\multirow{2}{*}{$V_2\sim(\bar 3,2,5/6)_V^{-1,-1}$} &$y\,\bar d^\dagger \bar\sigma^\mu V_{2\mu} L$ &
$\frac{c}{\Lambda^5} Q^\dagger \bar\sigma^\mu L V_{2\mu}^\dagger H \bar e L H$ &
$\mathcal{O}_{23}$ 
&3& $\frac{c\,y\,y_d\,y_l}{(4\pi)^6} \frac{v^2}{\Lambda}$ &
$M\lesssim 10^4$ GeV
\\

&$y\, Q \sigma^\mu V_{2\mu} \bar e^\dagger$&
$\frac{c}{\Lambda^5} Q^\dagger \bar\sigma^\mu L V_{2\mu}^\dagger H \bar e L H$ &
$\mathcal{O}_{44_{a,b,d}}$ 
&3& $\frac{c\,y\,g^2}{2(4\pi)^6} \frac{v^2}{\Lambda}$ &
$M\lesssim 10^{7}$ GeV 
\\

$ \tilde V_2\sim(\bar 3,2,-1/6)_V^{-1,-1}$
&$y\,\bar u^\dagger \bar\sigma^\mu \tilde V_{2\mu} L$ 
& $\frac{c}{\Lambda} Q^\dagger \bar \sigma^\mu LH \tilde V^\dagger_{2\mu}$
& $\mathcal{O}_{4_a}$ 
& 1& $\frac{c\,y\,y_u}{(4\pi)^2}\,\frac{v^2}{\Lambda}$  &  $M\lesssim 10^{12}$ GeV \\

\multirow{2}{*}{$ U_1\sim(3,1,2/3)_V^{-1,1}$} &$y\,Q^\dagger \bar\sigma^\mu U_{1\mu} L$ & $\frac{c}{\Lambda} \bar u ^\dagger \bar \sigma^\mu  L H U^\dagger_{1\mu}$
					      & $\mathcal{O}_{4_a}$ 
					      &1&  $\frac{c\,y\,y_u}{(4\pi)^2}\,\frac{v^2}{\Lambda}$ 
					      &    $M\lesssim 10^{12}$ GeV \\

&$y\,\bar d \sigma^\mu U_{1\mu} \bar e^\dagger$ & $\frac{c}{\Lambda} \bar u ^\dagger \bar \sigma^\mu L H U^\dagger_{1\mu}$
& $\mathcal{O}_8$ 
&2& $\frac{c\,y\,y_u\,y_d\,y_l}{(4\pi)^4}\,\frac{v^2}{\Lambda}$  &   $M\lesssim 10^{7}$ GeV \\

$U_3\sim(3,3,2/3)_V^{-1,1}$&$y\,Q^\dagger \bar\sigma^\mu U_{3\mu} L$ & $\frac{c}{\Lambda} \bar u ^\dagger \bar \sigma^\mu L  U_{3\mu}^\dagger H$ 
			   & $\mathcal{O}_{4_a}$ 
			   & 1&$\frac{c\,y\,y_u}{(4\pi)^2}\,\frac{v^2}{\Lambda}$  & $M\lesssim 10^{12}$ GeV \\

$\tilde U_1\sim(3,1,5/3)_V^{-1,1}$&$y\,\bar u \sigma^\mu \bar e^\dagger \tilde U_{1\mu}$
				  & $\frac{c}{\Lambda^5} \bar u^\dagger \bar\sigma^\mu L H \tilde U_{1\mu}^\dagger \bar e L H$ 
				  & $\mathcal{O}_{46}$ 
				  & 3&$\frac{c\,y\,g^2}{2(4\pi)^6} \frac{v^2}{\Lambda}$ 
				  & $M \lesssim 10^{7}$ GeV
\\

\bottomrule
\end{tabular}
\caption{Particles with quantum numbers ${\rm (SU(3)_c, SU(2)_L, U(1)_Y)}_{P}^{\mathcal{L},3\mathcal{B}}$ that couple to SM leptons at the renormalizable level, where $P=F,S,V$ denotes whether it is a fermion, scalar or vector. Fermions are 2-component Weyl fermions. The corresponding Dirac partner is denoted by a bar on top of the same symbol. The interaction with leptons is shown in the second column. We do not show the SU(2) contractions. 
In order to obtain a conservative upper bound for the mass $M$, we choose the lowest-dimensional and simplest $\Delta \mathcal{L}=2$ operator (third column).
After integrating out the particle, the operator in the fourth column is generated. The operator naming convention follows the general classification of Babu and Leung~\cite{Babu:2001ex} together with $\mathcal{O}_a^{\prime n}\equiv\mathcal{O}_a (H^\dagger H)^n$.
The fifth column provides the lowest loop order at which neutrino mass is generated and the sixth column shows an estimate for it following Ref.~\cite{deGouvea:2007qla}. From perturbativity considerations, $c,y\lesssim\mathcal{O}(1)$, and using couplings to the third family, this translates into an upper bound on $M$ which is shown in the last column. $W$-bosons in the loop lead to a further suppression by $g^2/2\simeq 0.2$. }
	\label{tab:particles}
\end{table*}
As outlined in the introduction the generation of Majorana neutrino mass requires the introduction of at least one new particle which couples to leptons and the existence of a $\Delta \mathcal{L}=2$ operator. 
This allows to obtain a conservative upper limit for the mass of the lightest new particle by demanding that the atmospheric neutrino mass scale is generated.
In the estimate we use third generation SM Yukawa couplings and order one values for the new unknown couplings. In the case of a model with several new particles, our analysis applies to the lightest particle of the model which typically generates the largest contribution to neutrino mass. 

In Tab.~\ref{tab:particles} we list all possible particles with lepton number (first column) which couple to leptons at the renormalizable level. The first four particles induce neutrino mass at tree level via the well-known
seesaw mechanisms (type-I~\cite{Minkowski:1977sc,Yanagida:1979as,GellMann:1980vs,Mohapatra:1979ia,Glashow:1979nm},
type-II~\cite{Magg:1980ut,Schechter:1980gr,Cheng:1980qt,Lazarides:1980nt,Wetterich:1981bx,Mohapatra:1980yp}, type-III~\cite{Foot:1988aq}) and via the mixing $m \,\bar L_1 L$ of a new vector-like lepton doublet $L_1$ with the SM lepton doublet $L$. Notice that there is no symmetry that allows the new Weinberg-like operator $L_1 H L H$ and forbids the usual one. However, this contribution may be significant for $m/M \lesssim 1$, which induces large mixing with the SM leptons and is therefore constrained by measurements in the charged lepton sector. Notice that it in this scenario neutrino masses are generated at tree level with the particles of the usual seesaws as mediators and therefore two new particles are needed. Finally, $\bar\Sigma_1$ generates the SM operator $\mathcal{O}_1^{\prime1}$ and thus may generate neutrino masses at tree-level with four insertions of the SM Higgs vacuum expectation value, but the most conservative bound is obtained for neutrino masses generated one-loop order. The remaining particles generate neutrino masses radiatively. Note also that for $\bar N$ the renormalizable Yukawa with the Higgs field generates Dirac neutrino masses at tree level after electroweak spontaneous symmetry breaking. This is the only case where, if lepton number is imposed as an exact symmetry at the perturbative level, neutrinos will be massive Dirac particles.\footnote{This would not be the case of the vector-like states $\bar \Sigma_0$ and $\bar \Sigma_1$, where a combination of the SM neutrinos and the neutral states of the new multiplets remains massless.} For the rest of the states, if lepton number is conserved, neutrinos would remain massless to all orders.

The second column displays the renormalizable coupling of
the new particle and defines its lepton number. In order to obtain a conservative upper bound on $M$ 
(see below), we choose the lowest-dimensional and simplest $\Delta \mathcal{L}=2$ operator. This operator is listed in the third column.
In some sense, our approach is technically
equivalent to studying the simplest models for each type of particle and
deriving their upper bound. 
The fourth column (named BL) lists the odd-dimensional~\cite{deGouvea:2014lva,Kobach:2016ami} $\Delta \mathcal{L}=2$ operator 
which is generated after
integrating out the new particle. We follow the naming convention of Babu and Leung as provided in Refs.~\cite{Babu:2001ex,deGouvea:2007qla} and introduce the additional notation $\mathcal{O}_a^{\prime n}\equiv\mathcal{O}_a (H^\dagger H)^n$.
The loop order $\ell$ at which neutrino masses are generated is given in the fifth column. The sixth column provides an estimate for neutrino mass by closing off loops of SM particles, following Ref.~\cite{deGouvea:2007qla}: Each loop contributes $(4\pi)^{-2}$, chirality-flips are proportional to the SM Yukawa coupling, and $W$-bosons contribute $g^2/2$. The Weinberg operator is induced via matching at loop-level, with neutrino masses generated in the form of Eq.~\eqref{eq:mnueq}.
As we are interested in conservative upper limits, we neglect any additional suppression and set $\epsilon=1$.
The constraint on the atmospheric mass scale translates into an upper bound on $\Lambda$ and consequently on $M$, as the EFT requires $M\leq \Lambda$.\footnote{The upper bound also applies to $\Lambda$, which encodes the heavier (unspecified and model-dependent) particle/s involved in the violation of lepton number.} This bound is conservative and shown in the last column.
We note that the upper limits derived are applicable to \emph{all} models involving a particular particle, as long it is the lightest one, which is phenomenologically the most interesting possibility.  
In the cases where several SU(2) contractions in the $\Delta \mathcal{L}=2$ SM operators are possible we select the ones that yield the most conservative upper limit. 

The upper limits on the mass in Tab.~\ref{tab:particles} are robust, model-independent and conservative within our assumptions, but not necessarily the strongest possible bounds for a particular model, because there may be extra suppressions, as discussed above. The bounds span several orders of magnitude, in the range $[10^6,10^{15}]$ GeV. Limits for dominant couplings to the first two families are obtained by a simple rescaling. Relaxing the perturbativity conditions on the couplings pushes all bounds up.
Clearly, the most promising particle to search for is a doubly-charged scalar due to its low upper limit of $10^6$ GeV, followed by $X_2$, $R_2$, $\tilde S_1$, $V_2$, and $\tilde U_1$ with upper limits that are one order of magnitude weaker. The Zee-Babu model~\cite{Zee:1985id,Babu:1988ki} is the simplest model which contains the doubly-charged scalar. Its large electric charge further makes it a very interesting candidate for searches at colliders via its decays into same-sign leptons~\cite{Alcaide:2017dcx}.

\subsection{Relation to well-known models} \label{comparison}
The Zee model \cite{Zee:1980ai} includes both the singly-charged scalar $h$ and a new scalar doublet $\phi\sim(1, 2,1/2)_S$.  Writing the interactions as $L y Lh$, $L^\dagger y_l \bar e^\dagger H $, $L^\dagger y^{\prime} \bar e^\dagger \phi$ and $\mu h^\dagger H  \phi$, neutrino mass is generated at one loop and its largest value reads
$
m^{\rm max}_\nu \sim y^{\prime}\,\tfrac{\,y_\tau y\,}{(4\pi)^{2} }\,\tfrac{ \mu v^2}{M^2}\,,
$
where $M = \max(m_h,m_\phi)$. Our approach can be used for $m_h< m_{\phi} $ and our estimation is recovered for $c=y^\prime\,\mu/m_{\phi}$. Using order one couplings and $\mu \leq m_{\phi}$, the upper limit reads $m_h< m_{\phi}\lesssim 10^5$ TeV, and from the exact formula, we indeed find the same result (see also Ref.~\cite{Herrero-Garcia:2017xdu} for a numerical analysis of the model).

The Zee-Babu model~\cite{Zee:1985id,Babu:1988ki} (see Refs.~\cite{Babu:2002uu,McDonald:2003zj,AristizabalSierra:2006gb,Nebot:2007bc,Ohlsson:2009vk,Herrero-Garcia:2014hfa} for detailed studies of its phenomenology) contains a doubly-charged scalar $k$ and a new singly-charged $h$. The possible terms $\bar e^\dagger y \bar e^\dagger k$, 
	$L y^{\prime} L h$, $L^\dagger  y_l  \bar e^\dagger H$, and $\mu k h^\dagger h^\dagger$ generate the largest value of neutrino mass
$
m^{\rm max}_\nu \sim  y^{\prime2}\,\tfrac{y_\tau^2\,y\,}{(4\pi)^{4}} \tfrac{ \mu v^2}{M^2}\,,
$
where $M = \max(m_h,m_k)$. For $m_k<m_h$ our general estimation is recovered for $c=y^{\prime2}\mu/m_h$. Using order one couplings and $\mu \lesssim m_{h}$, which can be derived from naturalness and the absence of charge-breaking minima, our estimate results in $m_{k}<m_h\lesssim 3000$ TeV. If $h$ is the lighter state, then in general we can only derive $m_h< 10^5$ TeV (like in the Zee model). However, if based on some other theoretical argument or observational fact one also \emph{knew} that the largest neutrino mass is generated via the Zee-Babu model, then $m_k\lesssim 3000$ TeV and therefore also $m_{h}\lesssim 3000$ TeV. This is however not the case in general. Performing a numerical scan, one finds: $m_h, m_{k}\lesssim 300$ TeV~\cite{Ohlsson:2009vk,Herrero-Garcia:2014hfa}. 

This exemplifies that the upper bounds in Tab.~\ref{tab:particles} are robust (model-independent) and conservative, but not necessarily the strongest possible bounds for a particular model, because there can be extra suppressions which could even exclude the model, for instance if there is a small violation of lepton number. Our analysis serves to identify the most promising particles to search for on general grounds.

\subsection{Higgs naturalness}
Generally the new particles contribute to the Higgs mass $m_H$ and thus it is possible to obtain an upper bound on the mass of the new particle from demanding a low fine-tuning of $m_H$. We define the theory at the scale $\Lambda$ and estimate the leading log-enhanced contribution for each case.

\paragraph{Scalar particles} with electroweak charges and mass $M$ contribute to the Higgs mass via their Higgs portal coupling $\lambda$ at one-loop order, 
\begin{equation}
	\delta m_H^2 \simeq - \,\frac{ \lambda \,N_w N_c}{16\pi^2}\, M^2 \ln\left(\frac{M^2}{\Lambda^2}\right) \;,
\end{equation}
where $N_{c}$ [$N_w$] denotes the dimension of the SU(3) [SU(2)] representation.
Even if absent at tree level, $\lambda$ is generated at one-loop order by electroweak gauge boson loops, 
\begin{equation}
	\delta \lambda \simeq \frac{3(Y^2 g^{\prime 4} + C_2 g^4)}{32\pi^2}\,\ln\left(\frac{M^2}{\Lambda^2}\right)\;,
\end{equation}
with the SU(2) Casimir invariant $C_2$ and hypercharge $Y$. Thus naturalness poses a limit on the scalar mass
\begin{equation}
	M \left|\ln \frac{M}{\Lambda}\right|\lesssim \frac{16\pi^2 |\delta m^2_H|^{1/2}_{\rm max}}{\sqrt{6 N_c ( 3 D g^4 + N_w Y^2 g^{\prime 4} )}}\,,
\end{equation}
where $D$ is the SU(2) Dynkin index and $|\delta m^2_H|^{1/2}_{\rm max}$ is the maximum correction to the Higgs mass that is considered natural.
In the type-II seesaw model, the trilinear coupling
$\mu$ also contributes to the Higgs mass, $\delta m_{H}^{2} \simeq 12\mu^{2} \ln(M^2/\Lambda^2)/(16\pi^2)$~\cite{Chao:2006ye,Schmidt:2007nq}, which translates into an upper bound $\mu |\ln(M/\Lambda)|^{1/2}\lesssim \sqrt{2/3}\pi |\delta m_H^2|^{1/2}_{\rm max}$. A similar bound can be obtained in the Zee model~\cite{Herrero-Garcia:2017xdu}.

\paragraph{New fermions} with mass $M$ and Yukawa coupling $y$ contribute to the Higgs mass at one-loop order, 
\begin{equation}
	\delta m_H^2 \simeq  \frac{4 N_c C  |y|^2}{16\pi^2}\, M^2 \ln \left(\frac{M^2}{\Lambda^2}\right)\;,
\end{equation}
with $C=2$ for the electroweak triplets $\bar \Sigma_i$ and $C=1$ for the electroweak doublet and singlet fermions. Particles with electroweak charges also contribute at two-loop order to the Higgs mass, 
\begin{equation}
	\delta m_H^2 \simeq \frac{8\kappa\, N_c (3 D g^4 + N_w Y^2 g^{\prime 4})}{(16\pi^2)^2}\, M^2 \ln\left(\frac{M^2}{\Lambda^2}\right)
\end{equation}
with $\kappa=1(1/2)$ for Dirac (Majorana) fermions.
Thus naturalness demands the fermion masses to obey
\begin{align}
	M \left|\ln \frac{M}{\Lambda}\right|^{1/2}& \lesssim \frac{2\pi  |\delta m^2_H|^{1/2}_{\rm max}}{|y|\sqrt{2N_c C}}\;, 
	\\
	M  \left|\ln \frac{M}{\Lambda}\right|^{1/2} &\lesssim \frac{4\pi^2  |\delta m^2_H|^{1/2}_{\rm max}}{\sqrt{\kappa N_c (3Dg^4+ N_wY^2g^{\prime4})}}\;.
\end{align}
\paragraph{Vector bosons.} For models with vector bosons Higgs naturalness is model-dependent, because there are additional contributions depending on how their mass is generated.

\begin{figure*}[t!]\centering
\includegraphics[type=png,ext=.png,read=.png,width=0.8\linewidth]{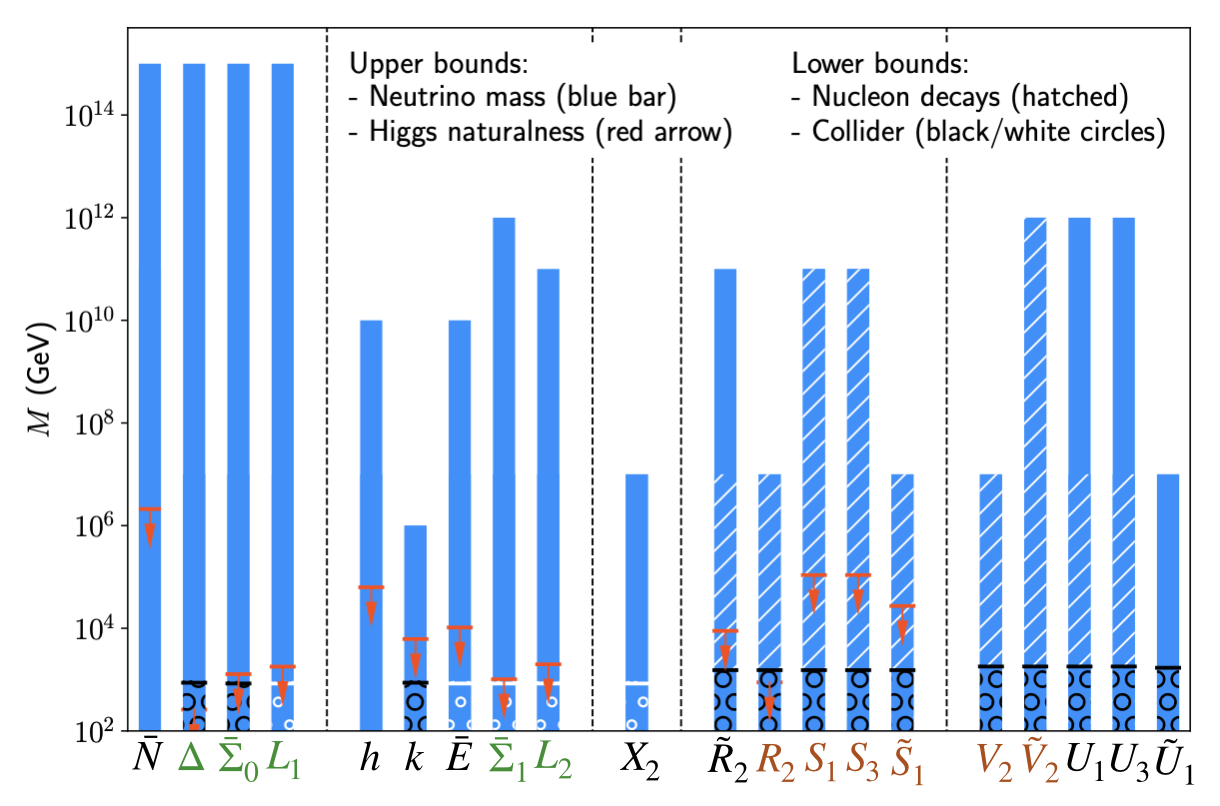}
	\caption{Summary plot of the upper limits. The blue bars illustrate the robust, model-independent and conservative upper limits from neutrino masses. If two upper limits are provided for a given particle in Tab.~\ref{tab:particles}, we use the most conservative one. The horizontal red lines indicate the upper limits from Higgs naturalness. Hatching indicates parameter space excluded by non-observation of $\mathcal{B}$-violating nucleon decays. In order to illustrate the current collider limits, we show current limits from ATLAS and CMS in black circles and estimated limits in white. These limits depend on the flavour structure and are thus model-dependent. We quote the most stringent lower limit. Particles that are excluded by combining the constraints from nucleon decays and neutrino masses are highlighted in red, and particles for which collider searches and Higgs naturalness limits are comparable in green.}
	\label{fig:summary}
\end{figure*}

In Fig.~\ref{fig:summary} we show the model-independent upper limits from neutrino mass as blue bars and indicate the upper limits from Higgs naturalness by horizontal red lines. We do not show masses below 100 GeV, because only a sterile neutrino $\bar N$ is allowed to be lighter. The renormalization scale is set to the maximally-allowed value of  $\Lambda$ from neutrino masses and $|\delta m^2_H|^{1/2}_{\rm max}=m_H=125$ GeV. The electroweak two-loop contribution generally dominates if present. For $\bar N$ there is only the one-loop contribution. In this case we use the neutrino mass scale to fix the Yukawa coupling $y$. The Higgs naturalness limits for the three seesaw models are consistent with previous results~\cite{Vissani:1997ys,Casas:2004gh,Farina:2013mla} taking the different choice of renormalization scale into account.

\section{Lower bounds\label{sec:lower}}
In this paper we do not attempt a complete study of the phenomenology, since it largely depends on the flavour structure. In the next subsection we illustrate how it is possible to use this framework to study it, while in the following subsections we make some general remarks.

\subsection{Studying flavour-dependent processes}
\label{sec:illustration}

The classification in terms of the lightest new particle and a $\Delta
\mathcal{L}=2$ effective operator can be used to study processes which do not
violate lepton number. We illustrate this using as an example the $S_1$ leptoquark with interaction terms $y_{ij} L_i Q_j S_1 + \tfrac{c_{ij}}{\Lambda}
L_i H S_1^\dagger \bar d_j$. If the contribution from the bottom quark
dominates and thus $(m_\nu)_{ij} = f [y_{i3} c_{j3}+ y_{j3} c_{i3}]/\Lambda$ with $f=f(m_b,m_{S_1})$, the
Yukawa coupling $y$ is determined in terms of neutrino masses and leptonic
mixing up to an overall unknown factor $\zeta$ and 2 discrete choices
($\pm$)~\cite{Cai:2014kra,Hagedorn:2018spx}, e.g. for normal mass ordering $y_{i3}=
\zeta^\pm \sqrt{\Lambda/(2 f)} (\sqrt{m_2}u_2^* \pm \sqrt{m_3}u_3^*) $, where $m_i$ are the
neutrino masses and $u_i$ the columns of the leptonic mixing matrix (Pontecorvo-Maki-Nakagawa-Sakata (PMNS) matrix). Similarly to Ref.~\cite{Cai:2014kra}, this determines i) the
branching ratios of $S_1\to b\nu_i, t\ell_i$ and thus provides a clear
prediction for collider searches;\footnote{The decays via the effective operator are generally suppressed.} ii) the relative
branching ratios for different processes, which are completely fixed. On the other hand the overall rate of lepton-flavour-violating
observables depends on the unknown combination of parameters $\zeta^\pm \sqrt{\Lambda/(2 f)}$. For more
complicated flavour structures it may be useful to use the recently-proposed
parametrisation of the neutrino mass matrix~\cite{Cordero-Carrion:2018xre}. A
detailed study of the phenomenology is left for future work.

\subsection{Charged lepton flavour/universality violation}
In the type-I seesaw model charged lepton flavour violation is suppressed due to the large scale of the new particles and unitarity (GIM mechanism).
The doubly-charged scalars $\Delta^{++}$ and $k$ induce tri-lepton decays at tree level and pose a stringent constraint on the involved Yukawa couplings $y/M \lesssim (g/m_W) [BR_{\rm lim}(l\to l_1l_2 \bar l_3)/BR(l\to l^\prime \nu\bar\nu^\prime)]^{1/4}$ in terms of the branching ratios $BR$ and the current limit $BR_{\rm lim}$. For instance, $\mathrm{BR}(\mu^{-}\rightarrow3e)<10^{-12}$ implies that the symmetric couplings of $k$ should satisfy $|y_{e\mu}y_{\mu\mu}^{*}|<2.3\cdot10^{-5}\,\left(\frac{m_{k}}{\mathrm{TeV}} \right)^{2}$.

First-generation leptoquarks may induce $\mu-e$ conversion at tree level and thus 
$y/M \lesssim (BR_{\rm lim} \,\omega_{\rm capt}/ (4\,C_N))^{1/4}$ where $\omega_{\rm capt}$ denotes the capture rate and $C_N\sim (0.01-0.1)\, m_\mu^5$ parameterizes the nuclear physics.
The contribution to radiative leptonic muon and tau decays can be estimated by
$BR(l\to l^\prime\gamma)/BR(l\to l^\prime\nu\bar\nu^\prime) \sim 3\alpha_{\rm em}y^4/(16\pi G_F^2 M^4)$, but may be further enhanced if the fermion in the loop is heavier than the decaying lepton.  For example, in the Zee-Babu model, $\mu\rightarrow e\gamma$ limits imply~\cite{Herrero-Garcia:2014hfa}
\begin{equation}
\frac{|y_{e\tau}^{\prime*}y^\prime_{\tau\mu}|^{2}}{(\frac{m_{h}}{\mathrm{TeV}})^4}+16 \frac{|y_{ee}^{*}y_{e\mu}+y_{e\mu}^{*}y_{\mu\mu}+y_{e\tau}^{*}y_{\tau\mu}|^{2}}{(\frac{m_{k}}{\mathrm{TeV}})^4}\lesssim 10^{-6}\,.
\end{equation}
A singly-charged scalar also generates violations of universality. In particular, the extracted Fermi constant from muon decay changes with respect to the SM value. Using the limits of the unitarity of the CKM, one obtains that its antisymmetric couplings should obey $|y^\prime_{e\mu}|^{2}<0.007\,\left(\frac{m_{h}}{\mathrm{TeV}} \right)^{2}$, and comparing different decay channels, one gets for example $||y^\prime_{e\tau}|^{2}-|y^\prime_{e\mu}|^{2}|<0.035\,\left(\frac{m_{h}}{\mathrm{TeV}} \right)^{2}$~\cite{Herrero-Garcia:2014hfa}.

Mixing with SM leptons leads to a non-unitary PMNS matrix~\cite{Antusch:2006vwa,Abada:2007ux,Abada:2008ea,He:2009tf,Fernandez-Martinez:2016lgt,Herrero-Garcia:2016uab}.
For the type-I and type-III seesaw models, it is generally small due to its relation to neutrino masses. However in extended models like
the inverse seesaw model or in models with new fermions with weak charges
$(\bar E,  L_i)$, there may be large deviations (See e.g. Ref.~\cite{Herrero-Garcia:2016uab}). The constraints of non-unitarity are typically $|y|^2/(\frac{M}{v})^2 \lesssim 10^{-3}$ (except for the first-second entry, where $\mu \rightarrow e \gamma$ implies the stronger constraint $10^{-5}$).

Generally the new particles also lead to non-standard interactions (NSI), see e.g.~Ref.~\cite{Babu:2019mfe} for a recent study in which it is found that significant NSI may still be allowed in some regions of the parameter space of radiative neutrino mass models.For example in the Zee model, the maximum NSIs depend on the assumptions made regarding the allowed fine-tuning of the off-diagonal Yukawas of the second Higgs doublet, which on the generic Higgs basis contribute to charged lepton masses, and also on the assumptions made on the trilinear $\mu$ term, which generates a correction to the Higgs mass at one loop.

\subsection{Lepton number violation}
The new particles in Tab.~\ref{tab:particles} may generate new contributions to $\Delta \mathcal{L}=2 $ probes, like neutrinoless double beta decay. The new contributions may be significant in some cases~\cite{Rodejohann:2011mu,Bonnet:2012kh}: (1) for type-I/III seesaw, if the new fermions have masses of order  $\mathcal{O}(1)$ GeV~\cite{Ibarra:2010xw,Blennow:2010th,Mitra:2011qr,Lopez-Pavon:2015cga}; (ii) if the scale $\Lambda$ of the relevant dimension-7 $\Delta \mathcal{L}=2$ operator, like $\mathcal{O}_8$, is low enough, $\Lambda\lesssim \mathcal{O}(100)$ TeV~\cite{deGouvea:2007qla}.\footnote{From neutrino masses, the scale of $\mathcal{O}_8$ is below $10^4$ TeV~\cite{deGouvea:2007qla}. $\mathcal{O}_8$ can be generated by $L_1,\,X_2,\,S_1,\,U_1$.} For type-II seesaw, the amplitude is very suppressed, by $(q/m_\Delta)^2$, where $q \sim 100$ MeV.

Another constraint comes from the fact that in the early universe sphaleron processes (active for temperatures $10^{12}\,{\rm GeV}\gtrsim T\gtrsim 100\,{\rm GeV}$) together with processes mediated by a $\Delta(\mathcal{B-L})=2$ operator may erase any previously-generated baryon asymmetry~\cite{Harvey:1990qw}. This imposes lower bounds on the scale
$\Lambda$ due to interactions mediated by either (i) the BL operators (fourth
column) or (ii) the $\Delta \mathcal{L}=2$ operator (third column),
if the new particle is relativistic and a
$\Delta \mathcal{L}=0$ interaction (e.g.~gauge interactions and/or the Yukawa coupling shown in the second column) rate is faster than the Hubble rate.
In particular, in order for a $\mathcal{B-L}$ asymmetry generated at $T_{\mathcal{B-L}}$ not
to be washed-out, for
order one couplings the requirement reads 
\begin{equation}
\Lambda \gtrsim [M_p\,T_{\mathcal{B-L}}^{2d-9}/(20 \mathrm{PS}_n)\,
]^{1/(2d-8)}\,,
\end{equation}
where  $M_p$ is the Planck scale, $d$ is the dimension of the operator (third and/or
fourth column) and PS$_n$ denotes the $n$-particle phase space factor.
For example, for two massless final state particles and $T_{\mathcal{B-L}}= 10^{6}, 10^{10}, 10^{12}$ GeV,
this reads $\Lambda \gtrsim 10^{11}, 10^{13}, 10^{14}$ GeV for the Weinberg operator and roughly $\Lambda \gtrsim 10^{7}, 10^{10}, 10^{13}$ GeV for other operators of dimension $d\leq11$. 
Notice that a lower limit on $M$ can be derived for order one couplings combining the washout lower limit with upper limit on $\Lambda$ from neutrino masses, if the exact combination of powers of $\Lambda$ and $M$ is known.

\subsection{Baryon number violation} 
There are stringent limits on baryon-number-violating ($\mathcal{B-L}$ conserving) dimension-6 operators with first generation quarks (See e.g.~\cite{Nath:2006ut,Barr:2012xb,Babu:2012qy,Arnold:2012sd,Dorsner:2012nq,Dorsner:2016wpm}) due to nucleon decays~\cite{Weinberg:1979sa,Wilczek:1979hc,Abbott:1980zj} such as $p\to e^+\pi^0$, $p\to \bar \nu \pi^+$ and $n\to\bar\nu \pi^0$, 
whose lower limit on the lifetime is $\mathcal{O}(10^{33})$ y~\cite{Abe:2013lua,Miura:2016krn}. 
Unless $B$-conservation is imposed, the scalar leptoquarks $S_1$, $S_3$ and $\tilde S_1$ have diquark couplings like $S_1\,\bar d \bar u$, $S_{1,3}\, Q^\dagger Q^\dagger$, $\tilde S_{1}\,\bar u \bar u$ and thus induce nucleon decay. The vectors $V_{2}$ and $\tilde V_{2}$ also have diquark couplings $\bar u \bar \sigma^\mu V_{2\mu} Q^\dagger$ and $\bar d \sigma^\mu \tilde V_{2\mu} Q^\dagger$, respectively, which mediate nucleon decay together with the other couplings shown in Tab.~\ref{tab:particles}. For $\tilde S_{1}$, the antisymmetry of the coupling implies that the decay proceeds into three leptons via $W$-boson exchange~\cite{Arnold:2012sd}, suppressed by $V_{td}\, y_t$ if coupled to the top quark. The lower limits on the mass are $M\gtrsim 10^{16}\,(10^{11})$ GeV for $\mathcal{O}(1)$ couplings for one lepton (three leptons in the case of $\tilde S_1$) in the final state, which are in tension with the neutrino mass bounds. 

There are also diquark couplings for $R_2,\,\tilde R_2$, generated by the $\mathcal{B}+\mathcal{L}$ conserving dimension-5 operators, $\tilde R_2 Q H^\dagger Q/\Lambda^\prime$ and $H^\dagger R_2 \bar d^\dagger\bar d^\dagger/\Lambda^\prime$. Similarly, the vectors $U_1,\,U_3$ also generate operators like $\bar d^\dagger  \sigma_\mu H^\dagger Q U^\mu_{1,3}/\Lambda^\prime$.
Therefore $\tilde R_2, \,U_1,\,U_3$ induce nucleon decays such as $n\to \pi^+ e^-$~\cite{Weinberg:1980bf,Weldon:1980gi} with decay width $\Gamma(n\to \pi^+ e^-) \simeq  y^2 \Lambda_{\rm QCD}^5 v^2/(8\pi \Lambda^{\prime2} M^4)$, where $\Lambda_{\rm QCD}\sim 1$ GeV. Using $\Gamma(n\to \pi^- e^+)^{-1} \gtrsim 5.3\times 10^{33} y$~\cite{TheSuper-Kamiokande:2017tit} and $M \leq \Lambda^\prime$, we obtain a lower limit $\Lambda^\prime \geq 10^{11}$ GeV for order one couplings, again in tension with the neutrino mass bound. 
Alternatively the scale $\Lambda^{\prime}$ can be taken to be the Planck mass, which leads to the lower bound $M\gtrsim 10^7$ GeV~\cite{Barr:2012xb,Babu:2012vc}. In the case of $R_2$, the antisymmetry makes it decay predominantly via the channel $p \rightarrow K^+ \nu$~\cite{Dorsner:2016wpm}, and the bounds are similar to those above. In Fig.~\ref{fig:summary}, we highlight the particles that are excluded by combining the constraints from nucleon decays and neutrino masses using red labels.

Due to its large hypercharge $\tilde U_1$ will only mediate nucleon decays via $\mathcal{B}$-violating operators of dimension larger than 5, involving multiple mesons and leptons, and thus it is currently not constrained.

$\mathcal{B}$-violating processes may also wash out the baryon asymmetry of the Universe, but these are generally more strongly constrained by nucleon decays. 

\subsection{Direct searches}
In the following we quote results from direct searches at colliders, which generically assume 100\% branching ratio in the considered channel. In realistic neutrino mass models, the limits are generally weaker due to multiple possible decay channels and thus reduced branching ratios. In any case there are already stringent lower bounds on several of the considered particles. Searches for two like-sign charged leptons constrain doubly-charged scalars $(\Delta^{++}, k)$ to be heavier than $M\gtrsim 770-870$ GeV, depending on the flavor~\cite{Aaboud:2017qph}. Similar limits are expected for $X_2$. The different particles can be distinguished by the different chirality of the final state leptons (see e.g.  Ref.~\cite{delAguila:2013yaa}). A model-independent bound of $M\gtrsim 200-220$ GeV is obtained for $\Delta^{++}$ by searching for $W^+W^-$ in the final state \cite{Aaboud:2018qcu}. 
The constraint on the fermionic triplet of type-III seesaw ($\bar \Sigma_0$) is $M>840\, \mathrm{GeV}$ assuming equal branching ratio to all flavors~\cite{Sirunyan:2017qkz}. Similar limits are expected for vector-like leptons ($\bar E$, $\bar \Sigma_1$, $L_1$, $L_2$).

Also leptoquarks have been searched for at the LHC. Neutrino masses are
generically dominated by third generation couplings. Searches for pair production of scalar leptoquarks with two $b$-jets
and $e\mu$ ($\tau\tau$) final states put a lower bound on the mass of $M >
640\, \mathrm{GeV}$ \cite{Aad:2015caa} ($M>850\, \mathrm{GeV}$
\cite{Sirunyan:2017yrk}). These searches do not apply to the $S_1$ leptoquark,
because it does not couple down-type quarks to charged leptons. There are also
constraints from searches for two jets and electrons (muons) which lead to more
stringent constraints of $M > 1010\, \mathrm{GeV}$ \cite{Khachatryan:2015vaa}
($M>1530\, \mathrm{GeV}$ \cite{Sirunyan:2018ryt}). Constraints on vector leptoquarks are generally a factor of $\sqrt{3}$ more stringent due to the multiple polarizations of the vector leptoquark. In particular CMS searches for vector leptoquarks decaying to a quark and a neutrino (electron or muon) now constrain the mass of the leptoquark to be larger than $1.8$ TeV ($1.7$ TeV)~\cite{Morse:2317226}.
Singly-charged scalars $h$ are bounded to be $\gtrsim 100$ GeV
from LEP data.
Finally, there are no competitive constraints yet for sterile neutrinos $\bar N$. %

In Fig.~\ref{fig:summary} we show the most stringent lower bounds from LHC
searches using black circles. This typically demands couplings to first or
second generation. For some of the particles there are no dedicated searches
and therefore no published lower limits. For them, we use the lower bound of a
particle which would yield the same signal, and highlight these bounds with
white circles. We find that for $\Delta,\,R_2$ ($\bar \Sigma_0,\,L_1,\,\bar
\Sigma_1,\,L_2),$ direct searches are in tension with (comparable to) Higgs
naturalness limits.\footnote{$R_2$ is already excluded by nucleon decays and
neutrino masses, and it is shown with a red label.} These are highlighted with
green labels. We emphasise once more that these collider limits, as well as the
other lower bounds, are model-dependent.

\section{Conclusions\label{sec:conclusions}}
We have derived general robust upper bounds on the mass of new particles
contributing to neutrino masses. Our main results are summarised in
Fig.~\ref{fig:summary}. We have also compared our limits with those
from Higgs naturalness, which are much stronger, but less robust. 
The most promising particles to search for are new doubly-charged scalars with masses below $\mathcal{O}(10^6)$ GeV, followed by $X_2$ and $\tilde U_1$ which also have low upper limits and are unconstrained by nucleon decays. Limits from direct searches together with Higgs naturalness arguments disfavour $\Delta,\,\bar \Sigma_0,\,L_1,\,\bar \Sigma_1,\,L_2,$ and $R_2$.

The lower bounds are generally model-dependent. Among these, in the cases where nucleon decays are generated,
they provide the most stringent limits. Limits from nucleon decays (taking conservatively the operators to be Planck
scale suppressed) imply that $S_1$, $S_3$, $\tilde S_1$, $R_2$, $V_2$ and
$\tilde V_{2}$ can not be the dominant source for neutrino masses, unless
baryon-number conservation is imposed. The vector bosons $V_2$ and $\tilde V_2,\,U_1$ are
naturally present as gauge bosons in Grand Unified Theories, for example in the 24
of SU(5) and the 45 of SO(10) (and in the Pati-Salam model), respectively. For SU(5)/SO(10)
the typical scale is incompatible with the gauge bosons being the dominant
source of neutrino masses. 
Our limits are compatible with leptogenesis
(for seesaw models) and low-scale generation of the baryon asymmetry of the Universe.
In particular the Davidson-Ibarra bound~\cite{Davidson:2002qv} is readily
satisfied by our upper bound for the type-I seesaw model. If baryon number is
conserved, all particles are allowed to be at the TeV scale (up to limits from direct searches) and provide the
dominant contribution to neutrino masses. Let us emphasize once more that our upper limits are very conservative, and in many models the scale will be much lower due to several chirality suppressions and/or small couplings, and/or if there is a small violation of lepton number $\epsilon$ (see Eq.~\eqref{eq:mnueq}, and first paragraph of section \ref{sec:upper}), like for example in inverse seesaw models.

This work is intended to serve not only as an indication of the most promising particles to directly search for at colliders, but also as a simple way of organising the plethora of neutrino mass models in \emph{just} 20 categories,
which allows for an easier study of their phenomenology.
If nucleon decays are not suppressed or forbidden by the absence of couplings to first generation quarks, or by baryon number conservation, it reduces to only 14 allowed categories. Direct searches and Higgs naturalness may further disfavour up to 5 more particles, leaving a final count of 9 allowed categories.
Lastly, we would like to emphasise that in order to explore the whole \emph{model space}, new dedicated collider searches for some of the particles (e.g.~the ones with white circled regions in Fig.~\ref{fig:summary}) are needed. 

\vspace{3ex}

\begin{acknowledgements}
We are grateful to A.~Santamaria, A.~Romanino, R.~Volkas, Y.~Cai, S.~Petcov and A. Vicente for useful discussions. This work has been supported in part by the Australian Research Council.
\end{acknowledgements}

\bibliographystyle{spphys} 
\bibliography{LNparticles}

\end{document}